\documentclass[twocolumn,superscriptaddress,showpacs]{revtex4}

\usepackage{mathrsfs}
\usepackage{amsthm}
\usepackage{amsmath}
\usepackage{CJK}
\usepackage{graphicx,subfigure,epstopdf}
\usepackage{indentfirst}
\usepackage{fancyhdr}
\usepackage{bm}
\usepackage{esint}
\usepackage{color}
\usepackage{multirow}

\newcommand{\ER}{Erd\H{o}s-R\'{e}nyi}

\newcommand{\ts}{t_c}
\newcommand{\tL}{t_{\Delta}}
\newcommand{\PRL}{\text{PR}_{\text{max}}}
\newcommand{\PRS}{\text{PR}_{\text{min}}}

\newcommand{\pc[1]}{p_{#1}} 
\newcommand{\pd[1]}{p_{#1}}

\begin{document}
\title{
Unstable supercritical discontinuous percolation transitions} 
%
\author{Wei Chen}
\email{chenwei2012@ict.ac.cn}
\affiliation{Institute of Computing Technology, Chinese Academy of Sciences, Beijing, China}
\affiliation{School of Mathematical Sciences, Peking University, Beijing, China}
\affiliation{University of California, Davis, California 95616, USA}

\author{Xueqi Cheng}
\affiliation{Institute of Computing Technology, Chinese Academy of Sciences, Beijing, China}

\author{Zhiming Zheng}
\affiliation{Key Laboratory of Mathematics, Informatics and Behavioral Semantics, Ministry of Education, Beijing University of Aeronautics and Astronautics, 100191 Beijing, China}

\author{Ning Ning Chung}
\affiliation{Department of Physics, National University of Singapore, 117542, Singapore}

\author{Raissa M. D'Souza}
\email{raissa@cse.ucdavis.edu}
\affiliation{University of California, Davis, California 95616, USA}

\author{Jan Nagler}
\email{jan@nld.ds.mpg.de}
\affiliation{Max Planck Institute for Dynamics and Self-Organization (MPI DS), G\"{o}ttingen}
\affiliation{Institute for Nonlinear Dynamics, Faculty of Physics, University of G\"{o}ttingen, G\"{o}ttingen, Germany}

\begin{abstract}

The location and nature of the percolation transition in random networks is  a subject of intense interest. 
Recently, a series of graph evolution processes have been introduced that lead to discontinuous percolation transitions where the addition of a single edge causes  
the size of the largest component to exhibit a significant macroscopic jump in the thermodynamic limit. These processes can have additional exotic behaviors, such as displaying a ``Devil's staircase" of discrete jumps in the supercritical regime. 
Here we 
investigate whether the location of the largest jump coincides with the percolation threshold for
%
a range of processes, such as \ER~percolation, percolation via edge competition and via growth by overtaking.
We find that the largest jump asymptotically occurs at the percolation transition for \ER~and other processes exhibiting global continuity, including models exhibiting an `explosive' transition. 
However, for percolation processes exhibiting genuine discontinuities, the behavior is 
 substantially richer. 
In percolation models where the order parameter exhibits a staircase,
the largest discontinuity generically does not coincide with the percolation transition.
For the generalized Bohman-Frieze-Wormald model, 
it depends on the model parameter. 
Distinct parameter regimes well in the supercritical regime feature unstable discontinuous transitions---a novel and unexpected phenomenon in percolation.
We thus demonstrate that seemingly and genuinely discontinuous percolation transitions can involve a rich behavior in supercriticality,
a regime that has been largely ignored in percolation.

\end{abstract}


\maketitle

\section{Introduction}

Percolation is a pervasive concept \cite{StaufferPercBook}, which has applications in a wide variety of natural, technological and social systems~\cite{DrosselPRL1992,ParshaniandBulyrevandStanley2010,CallawayPRL2000,NewmanandWatts2002},
ranging from conductivity of composite materials~\cite{Sahimi,AndradePRE2000} and polymerization~\cite{ZiffPRL1982} to epidemic  spreading~\cite{Anderson1991,CmoorePRE2000,PastorPRL2001} and 
information diffusion~\cite{StrangARS1998,Lazarsfeld1944}.
Viewed in a network setting, once the density of edges 
exceeds a critical threshold, $p_c$, the system undergoes a sudden transition to global connectivity, where the size of the largest connected component transitions from microscopic to macroscopic in size. If, rather than edge density, we consider the impact of adding individual edges, we expect to observe the largest jump in size of the largest component at $p_c$.  


It is well known that the classic {\ER}  (ER)~\cite{ERPS} model of percolation undergoes a continuous {\em second-order}  phase transition during  link-addition \cite{bollobas2001}.  Here one starts from a collection of $N$ isolated nodes and edges are added uniformly at random, with the critical edge density, $p_c=0.5$. Instead of \ER, we can consider {\em competitive percolation} processes \cite{Timme2010}.  In a competitive process, rather than a single edge, a fixed number of edges (or nodes) are chosen uniformly at random, but only the edge that best fits some specified criteria is added to the graph. Competition between edges is typically referred to as an Achlioptas Process~\cite{EPScience}.

Achlioptas Processes (AP) can exhibit a very sharp {\em explosive} transition which appears discontinuous on any finite system \cite{EPScience}.   
In past years such sharp transitions have been demonstrated
for
scale-free networks~\cite{ChoPRL09,RadFortPRL09,RadicchiPRE2010}, square lattices~\cite{ziffPRL09,Rziffnew}, Bethe lattice~\cite{ChaeKimPRE2012}, 
directed networks~\cite{Squires} and more realistic systems~\cite{KimYunPRE2012,Makse2009,RKPanPRE2011}. 
Although strong numerical evidence suggests that many explosive AP are discontinuous~\cite{EPScience}, 
more recently it has been shown that the seeming discontinuity at the percolation transition point disappears in the
 thermodynamic limit \cite{daCostaArxiv,Timme2010,Manna,RiordanWarnke,Grassberger,LeeKimPark}.
However, this neither means that all AP are necessarily globally continuous nor that there are no genuine discontinuities during the first continuous emergence of a giant component.
In fact, a giant connected component can emerge in a series of infinitely many genuinely discontinuous jumps and 
the notion that explosive percolation is always continuous \cite{RiordanWarnke} is thus misleading \cite{JNaglerDiscontinuous}.

\begin{figure*}
\begin{center}
 \includegraphics[width=0.95\textwidth]{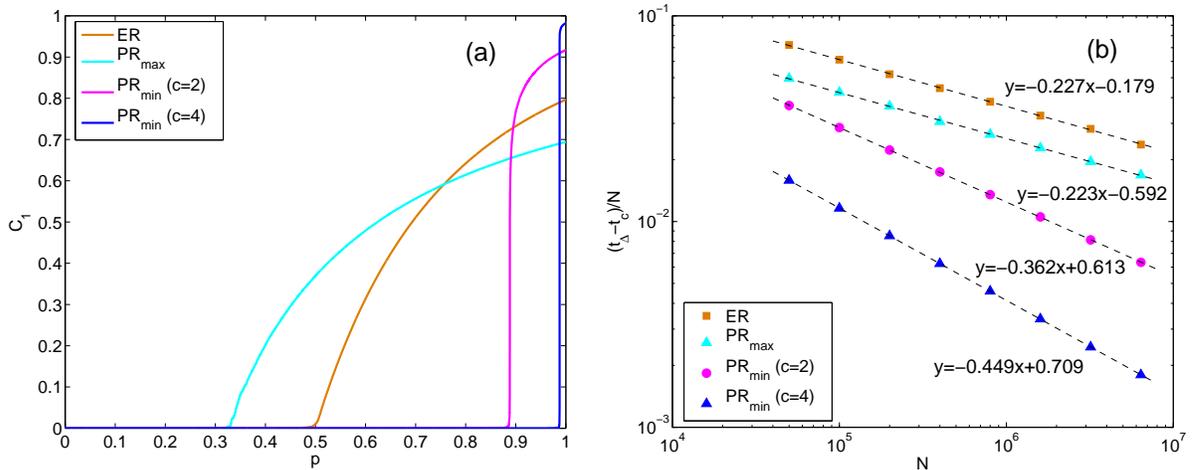}
 \end{center}
 \caption{{\bf Continuous percolation models.}
 (a) A typical evolution of $C_1$ as a function of the link density $p=t/N$ for ER, and for the $\PRL$, $\PRS$ models with the number of candidate edges $c=2$ and $c=4$, respectively. The system size $N=10^6$.
 (b) $(\tL-\ts)/N$ versus number of nodes $N$ for these models, 
 all of which follow a power-law distribution. 
 Each data point in (b) is averaged over $1000$ realizations.
 } 
 \label{fig:gs}
  \end{figure*}

Several random network percolation models have now been identified and studied that show a single genuine discontinuous transition~\cite{ChoKahng,Manna,Araujo,
WChenandRMDSouza,WChoi,Choarxiv,restrictedER,Boettcher2012,Bizhani2012,Caoarxiv,ChenNaglerDSouza2013,ChoScience2013}, or even multiple discontinuities~\cite{KJSchrenk,schrenk2012pre,JNaglerDiscontinuous}.
The mechanisms for discontinuous transitions such as
dominant overtaking~\cite{Timme2010,WChenandRMDSouzaarxiv}, cooperative phenomena~\cite{Bizhani2012} and the suppression principle~\cite{ChoPRL2012}, have received considerable attention, 
%
as have criteria to discriminate between continuous and discontinuous explosive percolation transitions. 
A signature for a continuous percolation transition is a critical power-law component-size distribution~\cite{RadicchiPRE2010, daCostaArxiv} 
and an asymptotically vanishing order parameter at the phase transition point~\cite{Araujo}. 


A method to
discriminate between weakly and genuinely discontinuous transitions proposed in~\cite{Timme2010} is to use the asymptotic size of the largest jump in the order parameter from the addition of a single edge. Importantly, if the largest jump of the order parameter does not vanish as the system size $N \rightarrow \infty$, 
 the transition is necessarily discontinuous. 
 However, whether the largest jump in the order parameter asymptotically coincides with the percolation transition point, and thus announces it, 
has remained largely unaddressed. 

In this paper, we study whether the position of the largest jump in the order parameter asymptotically converges to the percolation transition point.
To exemplify this we study
the {\ER} model, Achlioptas Processes and the generalized Bohman-Frieze-Wormald model (BFW)~\cite{TBohmanandAFrieze}. 
We find that the position of the largest jump in the order parameter asymptotically converges to
the percolation transition point for ER and AP with global continuity, but not necessarily for AP with discontinuities. 
For the BFW model, it
depends on the value of the model parameter $\alpha$. 
In BFW multiple giant components emerge at the percolation point, with the value of $\alpha$ determining the number of giants~\cite{WChenandRMDSouza}. 
Here we show that there are further sub-regimes of $\alpha$ values, with ``stable" regions where the macroscopic components never merge and ``unstable" regions where giants can have further merging in the supercritical regime.
In stable $\alpha$ regions the largest jump coincides with the percolation threshold, but in unstable regions the largest jump is in the supercritical regime. 

\section{Percolation models with global continuity}

We study whether the position of the largest jump in the order parameter asymptotically converges to the percolation transition point for percolation models exhibiting global continuity.
The best understood  percolation model that shows a continuous phase transition is the {\ER} model (ER)~\cite{ERPS}.
Let $N$ be the number of nodes, $t$ be the number of links (i.e., edges) in system and let 
 $C_i$ denote the fraction of nodes in the  $i$-th largest component. 
 A typical evolution of $C_1$, as a function of the link density $p=t/N$ (number of links per node) is shown in Fig.~\ref{fig:gs} (a). To study whether the position of the largest jump in $C_1$ converges to the percolation transition point, we measure $(\tL-\ts)/N$ 
as a function of $N$, where $\ts$  denotes the minimal number of steps for $C_1$ to exceed $N^{1/2}$,
and $\tL$ is the {\em time} (number of steps) when the largest jump in  $C_1$ has occurred. 
Fig.~\ref{fig:gs} shows that 
$(\tL-\ts)/N\sim N^{-0.227}$, which suggests that $(\tL-\ts)/N$ asymptotically converges to zero.
In addition
both $\tL/N$ and $\ts/N$ converge to the percolation transition point in the thermodynamic limit. 
%


\begin{figure*}
  \begin{center}
 \includegraphics[width=0.98\textwidth]{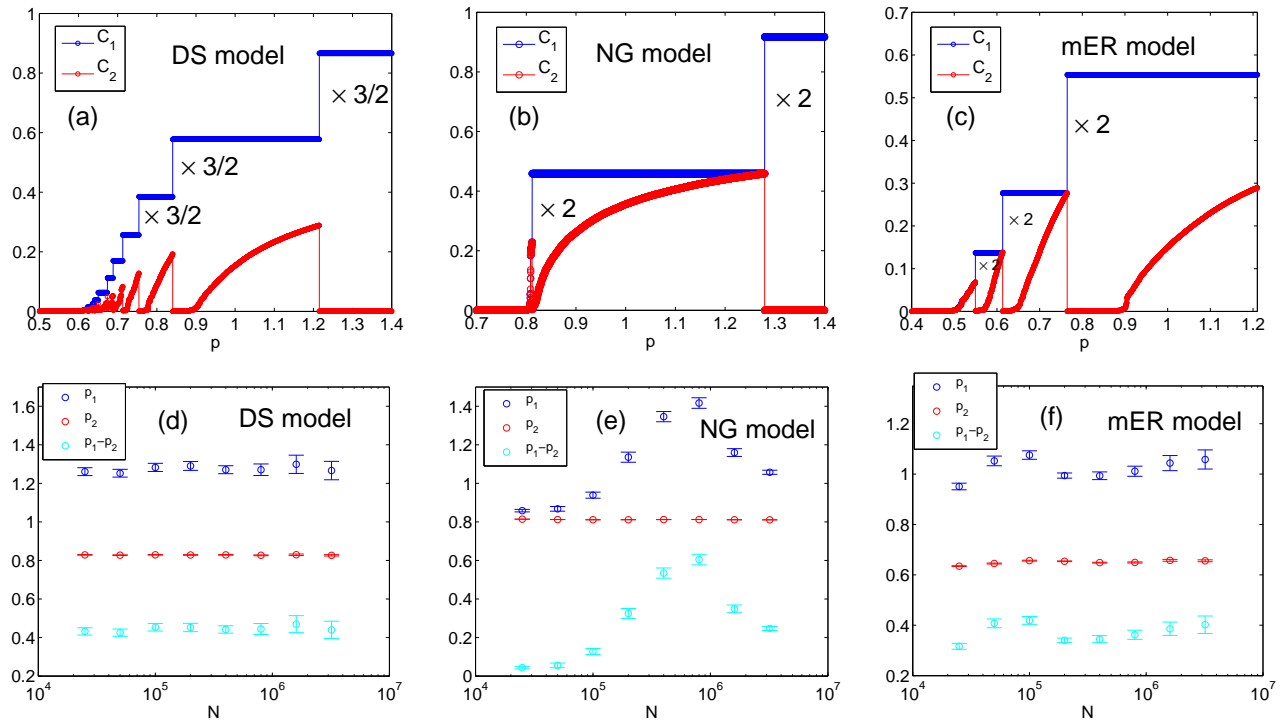}
 \end{center}
 \caption{{\bf Unstable supercritical discontinuous transitions.}
 Typical evolutions of $C_1$ and $C_2$ versus the link density for the DS model (a), NG model (b) and mER model (c) with system size $N=10^6$. 
 The link density $\pd[1]$ at which the largest jump in $C_1$ occurs, the link density $\pd[2]$ when the second largest jump in $C_1$ occurs and $\pd[1]-\pd[2]$ versus system size $N$ for the DS model (d), NG model (e) and mER model (f).
 Each data point in (d), (e) and (f) is averaged over $100$ realizations.  }
 \label{fig:hs}
  \end{figure*}

Next we study two extremal AP models, specifically
the $\PRL$ model and the `explosive' $\PRS$ model. 
In the $\PRL$ model, two candidate links are selected randomly at each step and the link that maximizes the product of the component sizes
that the ends of the link reside in is added while the other link is discarded. In the $\PRS$ model, $c$ candidate links are selected randomly at each step and the link that minimizes the product of the component sizes
that the ends of the link reside in is added while all other links are discarded. As an example, 
in Fig.~\ref{fig:gs}(a) we show the evolution of $C_1$ as a function of the link density $p=t/N$ for $\PRL$ and
 $\PRS$. 
 Similarily to the numerical results for the ER model, we find 
 that $(\tL-\ts)/N$ follows a power-law distribution for all studied models. 
 In particular $(\tL-\ts)/N\sim N^{-0.223}$ for $\PRL$, $\sim N^{-0.362}$ for $\PRS$ with $c=2$, and $\sim N^{-0.449}$ for  $\PRS$ with $c=4$. 
 
 This suggests that in continuous percolation, even if 
 extremely abrupt, the largest gap asymptotically announces the percolation transition.
This convergence, however, is not always guaranteed as we will show next.

\section{Models with Discontinuities}

In contrast to AP models discussed in the previous section, several random neighbor models (of the AP class) have been studied that
exhibit `staircase' discontinuities in the supercritical regime~\cite{JNaglerDiscontinuous,RiordanWarnkePRE}. Here we focus on three types of AP models, 
the Devil's staircase model (DS)~\cite{JNaglerDiscontinuous}, the Nagler-Gutch model (NG)~\cite{RiordanWarnkePRE},
 and the modified ER model (mER model)~\cite{RiordanWarnkePRE}. The DS model is based on picking three nodes at random and forbidding the largest picked component to merge with components whose sizes are not similar, 
 which results in a Devil's staircase with an infinite hierarchy of discontinuous jumps in $C_1$~\cite{JNaglerDiscontinuous}, see Fig.~\ref{fig:hs} (a).
This model has been used as a counterexample for the conclusion made in Ref. \cite{RiordanWarnke} that explosive percolation 
should be always continuous \cite{JNaglerDiscontinuous}.

Like the DS model, the NG model and mER model are both based on 3-node rules in which the addition of links connecting two components whose sizes are similar is favored. 
 Let $\pd[1]$ denote the link density immediately after the largest jump in $C_1$ from the addition of a single edge, and $\pd[2]$ denote the link density immediately after the second largest jump in $C_1$  from the addition of a single edge.
  Fig.~\ref{fig:hs} (d) shows that $\pd[1]$, $\pd[2]$ and $\pd[1]-\pd[2]$  asymptotically converge to some positive constant as $N\rightarrow\infty$, 
 because of the occurrence of multiple discontinuous jumps in the supercritical regime. 
 Figures~\ref{fig:hs} (b) and (c) show realizations of $C_1$ and $C_2$ for the NG and the mER model, in both cases
  featuring multiple discontinuous transitions of $C_1$~\cite{RiordanWarnkePRE}. 
  Figures~\ref{fig:hs} (e) and (f) show that  for both the NG model and the mER model, $\pd[1]$, $\pd[2]$ and $\pd[1]-\pd[2]$ asymptotically converge to some non-zero constant as the system size $N$ increases.
This numericial result can be understood in the following way.
From numerical observations, the size of the third largest component immediately before $\pd[2]$ in the supercritical regime is at most $\mathcal{O}(logN)$ for the DS model, the NG model and mER model.  
In addition, 
for all these models 
it has been analytically demonstrated that
 immediately before $\pd[2]$ both the size of the largest and second largest component are of order $\mathcal{O}(N)$, where
for the DS model it can be shown that $C_2=\frac{1}{2}C_1$, while for the NG model and the mER model $C_2=C_1$ \cite{RiordanWarnkePRE, JNaglerDiscontinuous}.
%
As a result, for these models once the largest and second largest component merge together inducing a discontinuous jump, the size of the second largest component drops to $\mathcal{O}(logN)$. 
Thus $\mathcal{O}(N)$ links are required before at $\pd[1]$ 
the second largest component grows again to size $\frac{1}{2}C_1$ (for the DS model), or to size $C_1$ for the NG model and the mER model, respectively.
Thus, $\pd[1]-\pd[2]$ is necessarily extensive and asymptotically converges to some positive non-zero constant.

This demonstrates the occurrence of multiple discontinuous transitions, including the transition with the largest discontinuity, in the supercritical regime and not at the percolation critical point as in traditional percolation.

In general, we use the term stable coexistence when all giant components emerging at the percolation transition point persist and remain separate throughout the supercritical regime. We use the term unstable coexistence when at least two giant components emerging at the percolation transition point merge together at some point in the supercritical regime. 
We find that all the models studied in this section display
unstable supercritical discontinuous transitions, which is a novel and unexpected feature in percolation. 
The model we study next shows even a quantitatively richer behavior, with some regions of stable coexistence and other regions of unstable coexistence. 


\section{Bohman-Frieze-Wormald Model}

\begin{figure*}
\begin{center}
 \includegraphics[width=0.8\textwidth]{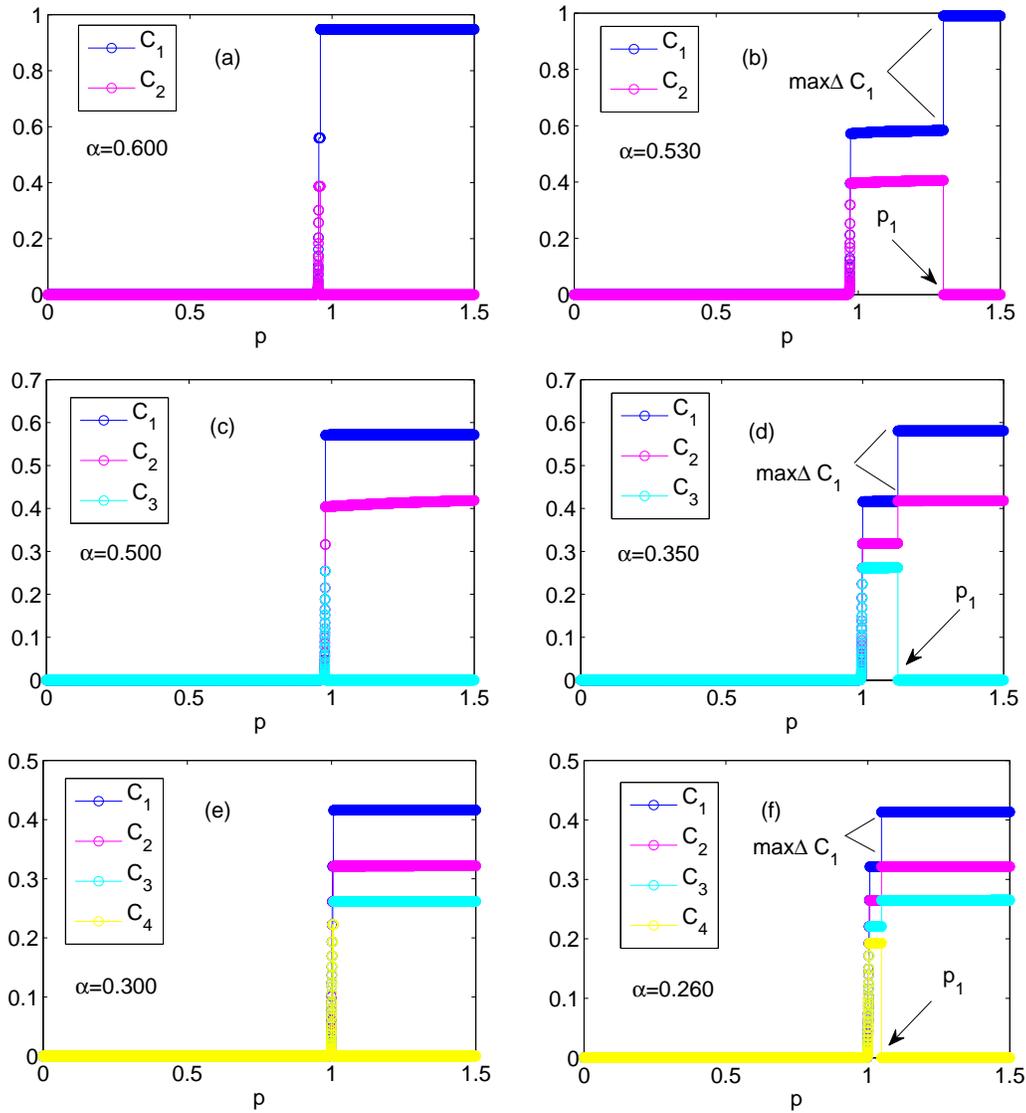}
 \end{center}
 \caption{ {\bf Critical and unstable supercritical discontinuous transitions.}
 (a) For $\alpha=0.600$, one giant component emerges in a phase transition. 
 (b) For $\alpha=0.530$, $C_1, C_2$ versus the link density, 
 showing that two giant components emerge simultaneously in the first phase transition.
 They are however unstable as they merge at a second transition. 
 (c) For $\alpha=0.500$,  $C_1, C_2, C_3$ versus the link density, showing that two giant components
 emerge simultaneously. 
 (d) For $\alpha=0.350$,  $C_1, C_2, C_3$ versus the link density, showing that three giant components emerge
 simultaneously in the first phase transition. 
 This configuration is unstable as in a second transition the second largest and the third largest components merge.
 (e) For $\alpha=0.300$, $C_1, C_2, C_3, C_4$ versus link density, showing the simultaneous emergence of three giant components. 
 (f) For $\alpha=0.260$,  $C_1, C_2, C_3, C_4$ versus density of links, showing that four giant components emerge simultaneously in the first
 phase transition. 
 Two smallest macroscopic components $C_3$ and $C_4$ merge together and overtake $C_1$, the other macroscopic components are stable in the remaining process, 
 $C_3+C_4\rightarrow C_1$, $C_1\rightarrow C_2, C_2 \rightarrow C_3$.
 System size is $N=10^6$ for all simulations.
 } 
 \label{fig:as}
  \end{figure*}
  
  \begin{figure*}
  \begin{center}
 \includegraphics[width=0.87\textwidth]{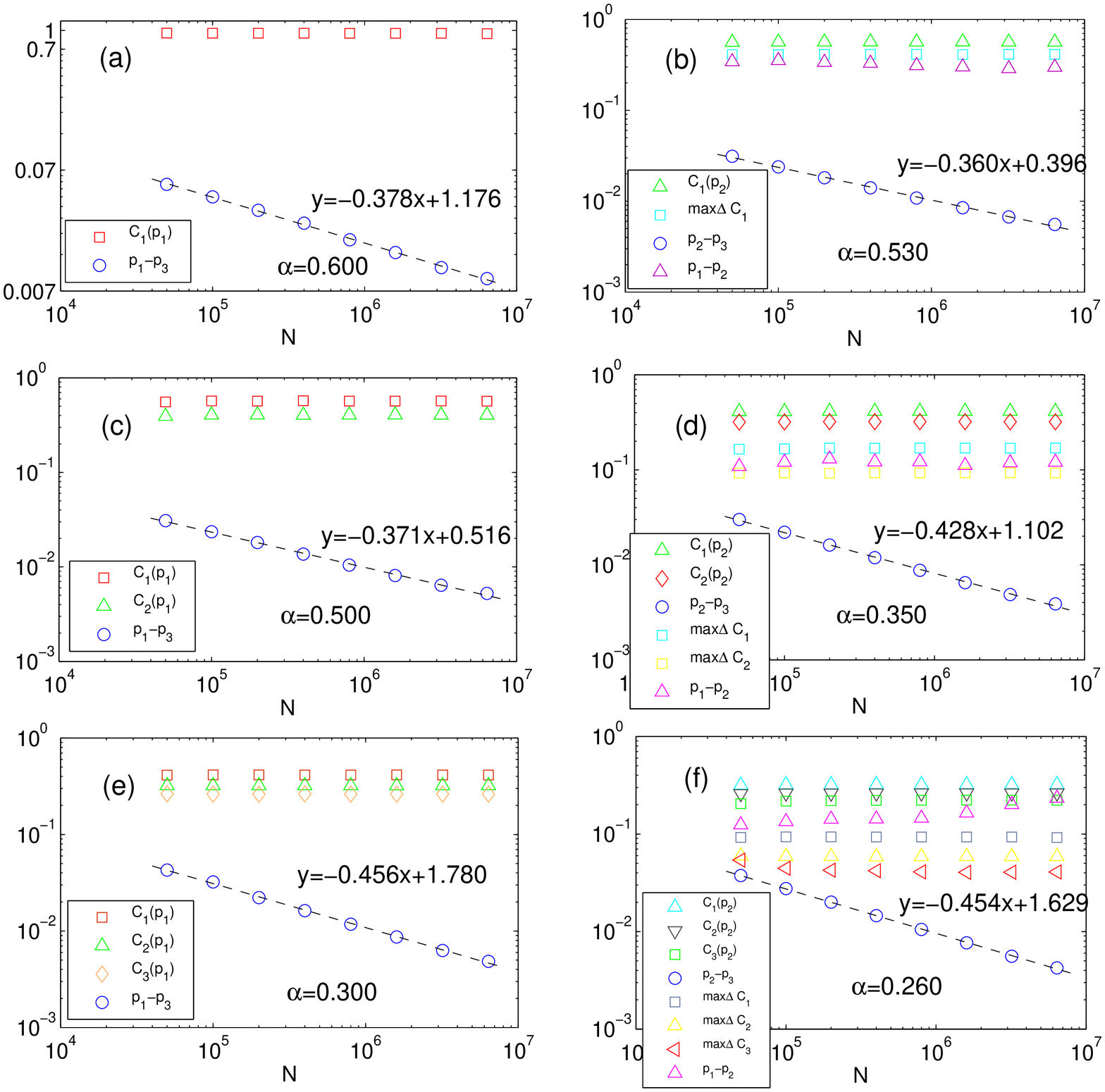}
 \end{center}
 \caption{{\bf Scaling of critical and unstable supercritical discontinuous transitions.}
 (a) For $\alpha=0.600$, $C_1$ at $\pc[1]$, $\pc[1]-\pc[3]$ versus system size. 
 (b) For $\alpha=0.530$, $C_1$ at $\pc[2]$, the largest jump in $C_1$, $\pc[2]-\pc[3]$, $\pc[1]-\pc[2]$ versus system size.
 (c) For $\alpha=0.500$, $C_1, C_2$ at $\pc[1]$, $\pc[1]-\pc[3]$ versus system size. 
 (d) For $\alpha=0.350$, $C_1, C_2$ at $\pc[2]$, the largest jump in $C_1, C_2$, $\pc[2]-\pc[3]$, $\pc[1]-\pc[2]$ versus system size. 
 (e) For $\alpha=0.300$, $C_1, C_2, C_3$ at $\pc[1]$, $\pc[1]-\pc[3]$ versus system size. 
 (f) For $\alpha=0.260$, $C_1, C_2, C_3$ at $\pc[2]$, the largest jump in $C_1, C_2, C_3$, $\pc[2]-\pc[3]$, $\pc[1]-\pc[2]$, versus system size. Each data point is averaged over $1000$ realizations.} 
 \label{fig:ms}
  \end{figure*}


The Bohman-Frieze-Wormald Model was first introduced 
to demonstrate that the emergence of a  giant component can be largely suppressed~\cite{TBohmanandAFrieze}.
(The details of BFW model can be found in the appendix.)
In particular, it has been established that under the BFW evolution, if $m=0.96689N$ links out of $2m$ sequentially sampled random links have been added to a graph, a giant component does not exist \cite{TBohmanandAFrieze}. 
More recently, the nature of the BFW transition was investigated. It was established that multiple giant components
appear in a discontinuous percolation transition for the BFW model~\cite{WChenandRMDSouza}.  In the asymptotic time limit of the BFW model, one-half of all edges that are sampled must be added.  Generalizing the BFW model by allowing the asymptotic fraction of edges to be a parameter $\alpha$, then the number of giant components can be tuned by adjusting the value of $\alpha$~\cite{WChenandRMDSouza}. 

For this model we investigate whether discontinuous jumps of the order parameter occur at the percolation transition point or in the supercritical regime.
First we study the evolution of the size of the four largest components, $C_i$ with $i=1,2,3,4$
for the BFW model with $\alpha\in(0,1]$,
as a function of the link density $p$. 

In Fig.~\ref{fig:as}(a) we show the discontinuous emergence of a unique giant component in a sharp transition to global connectivity, for $\alpha=0.6$.
For other values of $\alpha$, multiple giant components emerge simultaneously in a sharp transition. See, for instance, Fig.~\ref{fig:as}(c) for $\alpha=0.5$ with two giant components, or Fig.~\ref{fig:as}(e) for $\alpha=0.3$ with three giant components.
Thus, for certain values of $\alpha$, there exists a unique transition to global connectivity with multiple giant components.


Yet, Figs.~\ref{fig:as}(b), (d), (f) show there is another type of behavior possible, with an additional transition in the supercritical regime where giant components merge. This suggests the existence of an instability of giant components. 
Next, we perform numerical simulations to characterize the occurrence of discontinuities and multiplicities during the discontinuous transitions. 

As in the previous section, we denote the link density $\pc[1]$ as the position immediately after the largest jump in $C_1$ from the addition of a single edge, and $\pc[2]$ the position immediately after the second largest jump in $C_1$ from the addition of a single edge.  In addition, let $\pc[3]$ denote the minimal position at which the largest component contains at least $N^{1/2}$ nodes.
From numerical results in Fig.~\ref{fig:as}, we find that $\pc[1]\geq\pc[2]\geq\pc[3]$. 

Let us focus initially on the region with stable coexisting giant components. 
For $\alpha=0.6,0.5,0.3$, Fig.~\ref{fig:ms} (a), (c), (e) show that the size of the largest component at $\pc[1]$, denoted as $C_1(\pc[1])$, is almost independent of the system size $N$ and converges to some positive constant asymptotically. 
On the other hand, the gap between $\pc[1]$ and $\pc[3]$, i.e. $\Delta p=\pc[1]-\pc[3]$, scales as a negative power of $N$ and thus decreases to zero as $N\rightarrow\infty$, see Fig.~\ref{fig:ms}. 
This suggests that once the number of nodes in the largest component increases from $\mathcal{O}(N^{1/2})$ to $\mathcal{O}(N)$, the augmented link density converges to zero as $N\rightarrow\infty$, indicating the percolation process undergoes a unique discontinuous transition in the thermodynamic limit.

We further find that at $\pc[1]$, for $\alpha=0.5$, $C_2$ converges to some positive constant as well, see Fig.~\ref{fig:ms} (c),
and 
for $\alpha=0.3$, $C_1$, $C_2$, and $C_3$ all converge to some positive non-zero constant, respectively, see Fig.~\ref{fig:ms} (e).
This  indicates that the multiple giant components appear simultaneously in a unique discontinuous transition, consistent with the theory and observations put forth in~\cite{WChenandRMDSouza,WChenandRMDSouzaarxiv}.

Let us now focus on the regime with unstable coexistence of multiple giant components.
For $\alpha$ in the unstable regime, 
the size of the largest component at $\pc[2]$, 
is almost independent of the system size $N$ and converges to some positive constant asymptotically. 
For $\alpha= 0.53, 0.35, 0.26$, the size of the two, three and four largest components at $\pc[2]$ converge to some positive constant asymptotically. 
Yet, we find that $\pc[2]-\pc[3]$ decays as a power law in $N$, see Fig.~\ref{fig:ms} (b), (d), (f). This suggests the percolation transition is discontinuous at $\pc[2]$ with multiple giant components emerging simultaneously.

In addition, we observe that the size of the largest jump of the largest component at $\pc[1]$, 
denoted by $\max\Delta C_1$,
is independent of system size $N$ and converges to some positive constant asymptotically (see Fig.~\ref{fig:ms} (b), (d), (f)). 
This suggests the
occurrence of a second discontinuous transition at $\pc[1]$ in the supercritical regime.
We find that the second transition results from the merging of the two smallest giant components that emerge at the first (i.e., percolation) transition. 
This mechanism can be seen clearly from the case of $\alpha=0.26$ in Fig~\ref{fig:as} (f), where four giant components emerge at the first transition while at the second transition,
$\mathrm{C_3}$ and $\mathrm{C_4}$ merge together and overtake $\mathrm{C_1}$ in size. Thus $C_1, C_2, C_3$ all get a sudden jump in size but $C_4$ breaks down.
This overtaking mechanism dominates the growth of the largest component in the BFW model, which has been proven to be a key mechanism leading to discontinuous percolation transitions~\cite{WChenandRMDSouzaarxiv, Timme2010, JNaglerDiscontinuous}.

To test if the  transition points $\pc[1]$ and $\pc[2]$ in the unstable regime are still distinct in the thermodynamic limit, we next perform a scaling analysis.
For the values of $\alpha=0.53,0.35,0.26$, which are in the unstable region, we find 
that $\pc[1]-\pc[2]$ converges to a non-zero constant, 
see Fig.~\ref{fig:ms}.
This suggests, indeed, the distinctness of the two transition points. 

Taken together, we have identified and studied two parameter regimes in the BFM model,
(i) the stable regime of a unique discontinuous transition where one or more giant components emerge and coexist throughout the supercritical regime.
(ii) the unstable regime of multiple discontinuous transitions where multiple giant components emerge but the two smallest ones merge 
at a well-defined transition point in the supercritical regime. 

However, the characterization remains incomplete as so far we have only studied three instances for each parameter regime.
Our next aim is to establish a phase diagram by continuously tuning the parameter $\alpha$.


\begin{figure*}
\begin{center}
\includegraphics[width=0.7\textwidth]{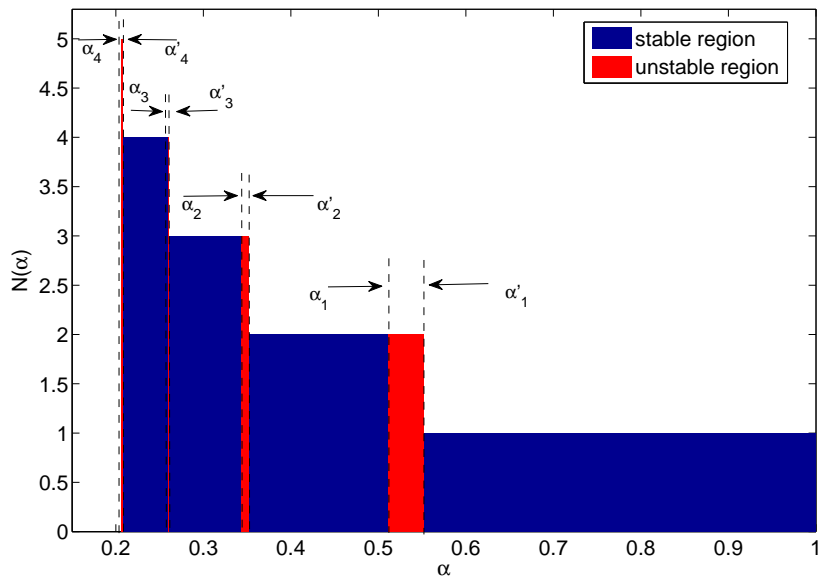}
\end{center}
\caption{{\bf Phase diagram for the BFW model.}
The number of giant components that appear in the first discontinuous phase transition (percolation transition point) in dependence on $\alpha$.
In the stable region, the giant components remain distinct in the supercritical regime while in the unstable region
the two smallest giant components merge in the supercritical regime at a well-defined transition point.
}
\label{fig:fs}
\end{figure*}

\subsection*{Phase Diagram of the BFW Model}
 
We first investigate the behavior in the regime 
$\alpha>\alpha_1$ with $\alpha_1= 0.511\pm0.003$,
where only one giant component asymptotically remains in the system. 
For values $\alpha<\alpha_1$, two giant components asymptotically remain~\cite{WChenandRMDSouza}.


Since in a stable regime of $\alpha$ all giant components that have emerged remain separate throughout the supercritical regime, a stable regime is characterized by  a unique transition of the largest component. 
In contrast, an unstable regime is characterized by
 one (or more) discontinuous transitions of $C_1$ in the supercritical regime by aggregation of two (or more) giant components emerging at percolation transition point. 

We find that the model undergoes two distinct discontinuous phase transitions for $\alpha\in(\alpha_1, \alpha'_1)$, referred to as the unstable regime, but undergoes a unique
discontinuous phase transition for $\alpha\in(\alpha'_1,1)$, referred to as the stable regime,  where $\alpha'_1=0.551\pm 0.001$. Previous work has established an infinite number bifurcation points $\alpha_i$,
at which the number of stable giant components that asymptotically remain in the system changes from $i$ to $i+1$, $i\geq 1$ (see table.~\ref{table1: summary})~\cite{WChenandRMDSouza}.
Here we numerically expand the analysis including transitions of the second largest component, the third largest component, and the forth largest component as well.
These transitions lead to multiple discontinuous transitions of the largest component and a hierarchy of stable and unstable regimes.
In particular, 
for $\alpha\in(\alpha'_{i},\alpha_{i-1})$, $i\geq2$, we identify a stable regime,
where in a unique discontinuous transition the $i$ largest macroscopic components, $C_1, C_2,... , C_i$, simultaneously emerge,
and for $\alpha\in(\alpha_i, \alpha'_{i})$ an unstable regime with two distinct discontinuous transitions, for $i\geq 1$. 
We find the numerical values $\alpha'_{2}=0.352\pm0.001, \alpha'_{3}=0.261\pm0.001, \alpha'_{4}=0.208\pm0.001...$, see table \ref{table1: summary}.

 
In Fig.~\ref{fig:fs} we demonstrate the phase diagram of the BFW model. 
We show the number of giant components that emerge at the first transition (percolation transition point) versus $\alpha$, with alternating stable and unstable regions. 
Since the number of giant components that asymptotically remain in the system increases as $\alpha$ decreases, there exist infinitely many bifurcation points $\alpha'_{i}, i \ge 1$, 
which separate stable from unstable regimes.

\begin{table}
\caption{\label{table1: summary}{\bf Summary of bifurcation points $\alpha_i$ and $\alpha'_i$.}
}
\centering{}
\begin{tabular}{lll}
\hline
\hline
$i$ & \ \ \ \ \ \ \ \ \ \ \ \ $\alpha_i$ & \ \ \ \ \ \ \ \ \ \ \ \ $\alpha'_{i}$\\
\hline
$1$ & \ \ \ \ $0.511\pm0.003$ & \ \ \ \ $0.551\pm0.001$\\
\hline
$2$ & \ \ \ \ $0.343\pm 0.001$ & \ \ \ \ $0.352\pm0.001$\\
\hline
$3$ & \ \ \ \ $0.259\pm0.001$ & \ \ \ \ $0.261\pm0.001$\\
\hline
$4$ & \ \ \ \ $0.206\pm0.001$ & \ \ \ \ $0.208\pm0.001$\\
\hline
\hline
\end{tabular}
\end{table}

\section{Summary and Discussions}

For various models with continuous, discontinuous and multiple-discontinuous percolation transitions,
we have investigated whether the location of the largest jump in the order parameter asymptotically converges to the percolation transition point marking the onset of global connectivity. 

For 
globally continuous transitions, including certain Achlioptas processes,
the location of the largest jump in the order parameter asymptotically converges to the percolation transition point. 
In contrast, Achlioptas processes with discontinuities exhibit a `staircase' of discontinuities in the supercritical region and the location of the largest jump is at an edge density well above the percolation point. 
Finally, the BFW model exhibits a rich supercritical behavior that is dependent on the model parameter $\alpha$,
as exemplified by the phase diagram, Fig.~\ref{fig:fs}, together with analytics suggesting an infinite hierarchy of regimes of alternating stability type.
Whether the percolation transition is asymptotically announced by the largest gap in the size of the largest component depends on the parameter. 
In the stable regime macroscopic components robustly coexist,
displaying the largest jump of the order parameter at the percolation transition point.
In the unstable region the coexistence of all macroscopic components that have emerged occurs in a finite sized window only, leading to multiple discontinuous transitions.
Macroscopic components that emerge at the percolation transition are thus not necessarily stable in the thermodynamic limit.

For AP models with discontinuities and for the BFW model, 
multiple discontinuous transitions are a consequence of the occurrence of extended periods in time well in the supercritical regime where macroscopic components cannot merge.
Mechanisms implying such periods are yet to be discovered. 

Multiple transitions have been studied in a wide variety of fields, such as geophysics~\cite{RamirezPRL2002}, liquid crystals~\cite{StrzeleckaNature}, classical thermodynamics and solid state physics~\cite{ArmstrongPRB2010}.
However, in random network percolation multiple transitions are poorly understood.
It would thus be interesting to identify the sufficient conditions for these \cite{JNaglerDiscontinuous, NaglerFractionalPercolation}.
This numerical work represents a step towards this direction.

In short, we have investigated unstable discontinuous transitions in percolation.
Seemingly and genuinely discontinuous percolation transition can involve a rich behavior in supercriticality, a regime that deserves attention for percolating systems
with substantial delays \cite{NaglerFractionalPercolation}.

\begin{acknowledgements}
This work was supported by the Defense Threat Reduction Agency, Basic Research Award No. HDTRA1-10-1-0088, the Army Research Laboratory under Cooperative Agreement Number W911NF-09-2-0053, the 973 National Basic Research Program of China (No. 2013CB329602, 2012CB316303), the National Natural Science Foundation of China (No. 11305219).
\end{acknowledgements}


\appendix*

\section{Algorithm of percolation models}

We state the algorithm of the percolation models studied in this paper in detail, which are the 
Devil's staircase model, the Nagler-Gutch model, the modified {\ER} model and the Bohman-Frieze-Wormald model.

\subsection{Devil's staircase (DS) Model}
Start with an empty graph with $N$ isolated nodes. 
At each step, three nodes $v_1, v_2, v_3$ are randomly selected from $N$ nodes, and let $s_1, s_2, s_3$ denote the sizes of components (not necessarily distinct) in which they reside.
Consider $\Delta_{i,j}=|s_i-s_j|$ with $1\leq i<j\leq 3$ and connect $v_i, v_j$ for which $\Delta_{i,j}$ is minimal.\\

\subsection{Nagler-Gutch (NG) Model}
Start with an empty graph with $N$ isolated nodes. 
At each step, three nodes $v_1, v_2, v_3$ are randomly selected from $N$ nodes, and let $s_1, s_2, s_3$ denote the sizes of components (not necessarily distinct) in which they reside.
Let $s_i$ denote the size of the component containing $v_i$. 
If all three component sizes $s_i$ are equal, add the edge connecting $v_1v_2$. If exactly two component sizes $s_i$ are equal, connect the corresponding nodes by a link. Otherwise
(if all $s_i$ are different), link the nodes in the two smallest components. This model is a modification of the ``explosive" triangle rule introduced in~\cite{RaissaMichael}, 
which in contrast exhibits a steep but continuous transition.\\

\subsection{Modified \ER \ (mER) Model}
Let $L_1$ and $L_2$ denote the sizes of the two largest components of the evolving graph. The mER model proceeds as follows. If the two largest components in the current graph have the same size ($L_1=L_2$), 
add the edge connecting $v_1v_2$. When $L_1>L_2$, if at least two $s_i$ are equal to $L_1$, connect two corresponding nodes, otherwise connect two nodes in components of size smaller than $L_1$.\\

\subsection{Bohman-Frieze-Wormald (BFW) Model}

Start with an empty graph with $N$ isolated nodes and proceed in phases starting with phase $k=2$. Edges are sampled one-at-a-time, uniformly at random from the complete graph. If an edge would lead to formation of a component of size less than or equal to $k$ it is accepted.  
Otherwise the edge is rejected provided that 
the fraction of accepted edges is greater than or equal to a function $g(k)$ that decreases with $k$. 
If the accepted fraction is not sufficiently large, the phase is augmented to $k+1$ repeatedly until either the edge can be accommodated or $g(k)$ decreases sufficiently that the edge can be rejected.  

Stated formally, let $k$ denote the stage, $N$ the number of nodes, $u$ denote the total number of links sampled and $A$ the set of accepted links (initially $A =\emptyset$), and $t=|A|$ the number of accepted links. 
At each step $u$, a link $e_{u}$ is sampled uniformly at random from the complete graph generated by the $N$ nodes, and the following algorithm iterated:  

{\tt Set $l=$ maximum size component in $A \cup \{e_{u}\}$

\ \ \ if $\left(l\leq k\right) \{$
 
\ \ \ \ \ \ $A \leftarrow A \cup \{e_{u}\}$
 
 \ \ \ \ \ \ $u\leftarrow u+1 \ \}$
  
\ \ \ else if $\left(t  / u < g(k)\right) \{$
 $k\leftarrow k+1\ \}$
 
\ \ \  else $\{\ u \leftarrow u+1 \ \}$

}
\noindent
where $g(k)=\alpha+\sqrt{1/2k}$ with $\alpha\in[0,1]$. Thus $\alpha$ denotes the asymptotic fraction of accepted links over totally sampled links. In the original BFW model, $\alpha=1/2$~\cite{TBohmanandAFrieze}. We generalize the BFW model by tuning $\alpha$.


\end{document}